\documentclass[a4paper,fleqn,usenatbib]{mnras}

\usepackage{newtxtext,newtxmath}
\usepackage[T1]{fontenc}
\usepackage{ae,aecompl}

\usepackage{graphicx}	% Including figure files
\usepackage{amsmath}	% Advanced maths commands
\usepackage{amssymb}	% Extra maths symbols
\usepackage{deluxetable}

\title[Stellar Angular Diameters from Photometry]{Predicting Stellar Angular Diameters from $V$, $I_C$, $H$, $K$ Photometry}

\author[A. D. Adams et al.]{
Arthur D. Adams,$^{1}$\thanks{E-mail: arthur.adams@yale.edu}
Tabetha S. Boyajian,$^{1,2}$
Kaspar von Braun$^{3}$
\\
$^{1}$Department of Astronomy, Yale University, 52 Hillhouse Avenue, New Haven, CT 06511, USA\\
$^{2}$Department of Physics and Astronomy, Louisiana State University, Nicholson Hall, Tower Drive, Baton Rouge, LA 70803, USA\\
$^{3}$Lowell Observatory, 1400 W. Mars Hill Road, Flagstaff, AZ 86001, USA
}

\date{Accepted XXX. Received YYY; in original form ZZZ}

\pubyear{2017}

\begin{document}
\label{firstpage}
\pagerange{\pageref{firstpage}--\pageref{lastpage}}
\maketitle

\begin{abstract}
Determining the physical properties of microlensing events depends on having accurate angular sizes of the source star. Using long-baseline optical interferometry we are able to measure the angular sizes of nearby stars with uncertainties $\leq 2\%$. We present empirically derived relations of angular diameters that are calibrated using both a sample of dwarfs/subgiants and a sample of giant stars. These relations are functions of five color indices in the visible and near-infrared, and have uncertainties of 1.8--6.5\% depending on the color used. We find that a combined sample of both main-sequence and evolved stars of A--K spectral types is well fit by a single relation for each color considered. We find that in the colors considered, metallicity does not play a statistically significant role in predicting stellar size, leading to a means of predicting observed sizes of stars from color alone.
\end{abstract}


%%%%%%%%%%%%%%%%%%%%%%%%%%%%%%%%%%%%%%%%%%%%%%%%%%

%%%%%%%%%%%%%%%%% BODY OF PAPER %%%%%%%%%%%%%%%%%%

\section{Introduction}
\label{introduction} 
Precise stellar radius measurements are important for many subfields of astronomy, especially for exoplanet characterization. While precise radii are most readily applicable to transiting exoplanet characterization, they also correspond directly to stellar angular diameters. One notable application for such angular diameters is in constraining the physical properties of microlensing events, for example in distinguishing cases of self-lensing from those of MACHO lensing \citep{cal10, fuk15}.
 
Microlensing systems are often far too distant for direct measurements of the stellar angular size, prompting empirical means to determine stellar sizes from photometry alone.

The surface brightness of a star for a given magnitude is defined in terms of the magnitude and angular diameter \citep{wes69,bar76,dib05}:
\begin{equation}
S_V = V_0 + 5 \log \theta
\end{equation}
where $V_0$ is an intrinsic magnitude set such that $S_V = V_0$ when the angular diameter $\theta = 1\,\mathrm{mas}$.

\citet{wes69} demonstrates a strong empirical correlation between surface brightness and $\left( B - V \right)$ color; a more general correlation between surface brightnesses and color indices has been shown in \citet{bar76}. Therefore we expect to be able to construct relations between stellar angular size, color, and an apparent magnitude from the given color. \citet{bar76} further demonstrate that surface brightness is independent of stellar luminosity class, which implies that such an angular size-color-magnitude relation should hold regardless of whether stars have evolved off the main sequence. \citet{dib05} proposed, through comparison of their empirical relations for both dwarf and giant stars, that there was enough overlap in the then available data to motivate a combined fit across evolutionary stages.

One photometric magnitude of each color is used as a baseline for developing a \emph{zero-magnitude diameter}, the angular diameter each star would appear to have if its apparent magnitude were zero in a selected band:
\begin{equation}
\log \theta_{Q=0} = \log \theta_{LD} + 0.2 Q
\end{equation}
where $\theta_{LD}$ is the angular diameter after correction for limb-darkening and $Q$ is the magnitude in a given band. We construct our relations as polynomials in color. For a given color $\left(P - Q\right)$
\begin{equation}\label{eq:polyfit}
\log \theta_{Q=0} = \sum_{n=0}^N c_n \left(P - Q\right)^n
\end{equation}
where $N$ is an arbitrary order, taken to be the greatest statistically significant order when fitting the data. Determination of angular sizes from observed colors is insensitive to wavelength-dependent extinction for the precisions attainable through this analysis \citep{bar76}; therefore we neglect extinction correction.

 The use of interferometry to measure the angular diameters of stars has played a major role in empirically constraining the radii of nearby stars. Our new relations benefit from recent precise angular diameter measurements of both main-sequence and evolved stars using optical interferometry. They extend the results of \citet{boy14} with new data and more precise relations for both main-sequence and evolved stars, constructed for a more limited range of spectral types.

Section \ref{data} describes the criteria for data selection and sources of angular diameters and photometry, and the methodology for fitting the data is presented in Section \ref{fitting}. We analyze the results in Section \ref{analysis}, including a comparison with previous works (Section \ref{sub:comp}).

\section{Data}
\label{data}

We compile a list of stars with both $V$, $I_C$, $H$, and/or $K$ magnitudes and precise angular diameters in Tables \ref{table:dwarfs_table}--\ref{table:giants_table}. A total of 57 distinct main-sequence stars are selected among all relations, with effective temperatures of 3927--9553 K (spectral types A1--M0), a mean angular diameter uncertainty of 0.013 mas, and apparent $V$ magnitudes of 0.03--7.70. The evolved sample contains 50 stars with effective temperatures of 3972--10330 K (spectral types A1--M0), a mean angular diameter uncertainty of 0.043 mas, and apparent $V$ magnitudes of 1.16--6.18. The following subsections outline both the source information as well as selection and classification criteria.

\begin{table*}
\centering
\caption{Selected Stellar Properties -- Dwarfs}
\label{table:dwarfs_table}
\resizebox{0.85\textwidth}{!}{
\begin{tabular}{|cccccccc|}
\hline
HIP & Sp. Type & $\theta_{LD}$ (mas) & $V$ & $I_C$ & $H$ & $K$ & $\theta_{LD}$ Ref. \\
\hline
3765  &  K2.5V  &  $0.868\pm0.004$  &  5.740  &  $4.780\pm0.027$  &  --  &  --  & 1 \\
3821  &  F9V  &  $1.623\pm0.004$  &  3.460  &  --  &  $2.020\pm0.050$  &  $1.821\pm0.060$  & 2 \\
4151  &  F9V  &  $0.865\pm0.010$  &  4.800  &  $4.210\pm0.027$  &  $3.560\pm0.050$  &  --  & 2 \\
4436  &  A6V  &  $0.708\pm0.013$  &  3.860  &  $3.700\pm0.027$  &  $3.370\pm0.060$  &  $3.365\pm0.071$  & 3 \\
5336  &  K1V Fe-2  &  $0.972\pm0.009$  &  5.170  &  $4.360\pm0.027$  &  --  &  --  & 4 \\
7513  &  F9V  &  $1.143\pm0.010$  &  4.100  &  $3.500\pm0.027$  &  $2.990\pm0.050$  &  $2.841\pm0.080$  & 5, 6 \\
7981  &  K1V  &  $1.000\pm0.004$  &  5.240  &  $4.360\pm0.027$  &  $3.345\pm0.050$  &  --  & 7 \\
8102  &  G8.5V  &  $2.080\pm0.030$  &  3.490  &  $2.630\pm0.027$  &  $1.727\pm0.050$  &  $1.631\pm0.060$  & 8 \\
12114  &  K3V  &  $1.030\pm0.007$  &  5.790  &  $4.740\pm0.027$  &  $3.542\pm0.050$  &  --  & 1 \\
12777  &  F7V  &  $1.103\pm0.009$  &  4.100  &  $3.530\pm0.027$  &  $3.070\pm0.050$  &  $2.761\pm0.090$  & 2 \\
16537  &  K2V (k)  &  $2.126\pm0.014$  &  3.720  &  --  &  $1.749\pm0.050$  &  $1.601\pm0.060$  & 9 \\
16852  &  F9IV-V  &  $1.081\pm0.014$  &  4.290  &  $3.640\pm0.027$  &  --  &  $2.871\pm0.100$  & 2 \\
19849  &  K0.5V  &  $1.446\pm0.022$  &  4.430  &  $3.530\pm0.027$  &  --  &  --  & 1, 10 \\
22449  &  F6IV-V  &  $1.419\pm0.027$  &  3.190  &  $2.650\pm0.027$  &  $2.148\pm0.050$  &  $2.031\pm0.060$  & 11, 2 \\
24813  &  G1V  &  $0.981\pm0.015$  &  4.690  &  $4.040\pm0.027$  &  $3.330\pm0.050$  &  $3.255\pm0.045$  & 2 \\
27435  &  G2V  &  $0.572\pm0.009$  &  5.970  &  --  &  $4.499\pm0.050$  &  --  & 7 \\
27913  &  G0V CH-0.3  &  $1.051\pm0.009$  &  4.390  &  --  &  $3.050\pm0.050$  &  $2.971\pm0.070$  & 2 \\
32349  &  A0mA1Va  &  $5.959\pm0.059$  &  -1.440  &  $-1.430\pm0.027$  &  $-1.387\pm0.050$  &  --  & 12, 13, 14, 15, 16 \\
32362  &  F5IV-V  &  $1.401\pm0.009$  &  3.350  &  $2.870\pm0.027$  &  --  &  $2.111\pm0.060$  & 2 \\
35350  &  A3V  &  $0.835\pm0.013$  &  3.580  &  $3.450\pm0.027$  &  --  &  --  & 2 \\
36366  &  F1V  &  $0.853\pm0.014$  &  4.160  &  $3.780\pm0.027$  &  --  &  --  & 2 \\
37279  &  F5IV-V  &  $5.434\pm0.050$  &  0.400  &  $-0.140\pm0.027$  &  $-0.569\pm0.050$  &  $-0.669\pm0.051$  & 13, 17, 18, 19 \\
40843  &  F6V  &  $0.706\pm0.013$  &  5.130  &  --  &  $3.940\pm0.050$  &  --  & 7 \\
43587  &  K0IV-V  &  $0.711\pm0.004$  &  5.960  &  --  &  $4.140\pm0.050$  &  --  & 20 \\
45343  &  M0.0V  &  $0.871\pm0.015$  &  7.640  &  --  &  $4.253\pm0.050$  &  --  & 1 \\
46733  &  F0V  &  $1.133\pm0.009$  &  3.650  &  $3.270\pm0.027$  &  --  &  $2.711\pm0.090$  & 2 \\
46853  &  F7V  &  $1.632\pm0.005$  &  3.170  &  $2.610\pm0.027$  &  $2.025\pm0.050$  &  $1.951\pm0.070$  & 2 \\
47080  &  G8IV  &  $0.821\pm0.013$  &  5.400  &  --  &  $3.770\pm0.050$  &  --  & 2 \\
51459  &  F8V  &  $0.794\pm0.014$  &  4.820  &  $4.240\pm0.027$  &  --  &  --  & 2 \\
53910  &  A1IVspSr  &  $1.149\pm0.014$  &  2.340  &  $2.380\pm0.027$  &  --  &  $2.361\pm0.060$  & 2 \\
56997  &  G8V  &  $0.910\pm0.009$  &  5.310  &  $4.580\pm0.027$  &  --  &  --  & 2 \\
57757  &  F8.5IV-V  &  $1.431\pm0.006$  &  3.590  &  $3.000\pm0.027$  &  $2.345\pm0.050$  &  $2.301\pm0.060$  & 2 \\
57939  &  G8. V P  &  $0.686\pm0.006$  &  6.420  &  $5.570\pm0.027$  &  --  &  --  & 2, 21 \\
64394  &  G0V  &  $1.127\pm0.011$  &  4.230  &  $3.620\pm0.027$  &  $2.923\pm0.050$  &  $2.851\pm0.100$  & 2 \\
64924  &  G7V  &  $1.073\pm0.005$  &  4.740  &  $3.990\pm0.027$  &  --  &  --  & 22 \\
65721  &  G5V  &  $1.010\pm0.020$  &  4.970  &  $4.190\pm0.027$  &  $3.320\pm0.050$  &  --  & 5 \\
66249  &  A2Van  &  $0.852\pm0.009$  &  3.380  &  $3.280\pm0.027$  &  $3.050\pm0.050$  &  --  & 2 \\
67927  &  G0IV  &  $2.252\pm0.036$  &  2.680  &  $2.080\pm0.027$  &  $1.390\pm0.050$  &  $1.291\pm0.051$  & 13, 18, 23, 24 \\
71284  &  F4VkF2mF1  &  $0.841\pm0.013$  &  4.470  &  $4.020\pm0.027$  &  $3.516\pm0.050$  &  --  & 2 \\
72567  &  F9IV-V  &  $0.569\pm0.011$  &  5.860  &  --  &  $4.530\pm0.050$  &  --  & 7 \\
72659  &  G7V  &  $1.196\pm0.014$  &  4.540  &  --  &  $3.000\pm0.050$  &  $2.651\pm0.080$  & 2 \\
78459  &  G0V  &  $0.735\pm0.014$  &  5.390  &  --  &  $3.945\pm0.050$  &  $3.901\pm0.045$  & 22 \\
81300  &  K0V (k)  &  $0.724\pm0.011$  &  5.770  &  --  &  $3.910\pm0.050$  &  --  & 1 \\
91262  &  A1V  &  $3.280\pm0.010$  &  0.030  &  $0.080\pm0.027$  &  $0.004\pm0.050$  &  $-0.079\pm0.060$  & 13, 16, 25, 26, 27 \\
92043  &  F5.5IV-V  &  $1.000\pm0.006$  &  4.190  &  $3.660\pm0.027$  &  --  &  $2.941\pm0.090$  & 2 \\
93747  &  A1V  &  $0.895\pm0.017$  &  2.990  &  $2.990\pm0.027$  &  --  &  $2.921\pm0.080$  & 2 \\
96100  &  G9V  &  $1.254\pm0.012$  &  4.670  &  $3.850\pm0.027$  &  --  &  $2.811\pm0.080$  & 4 \\
96441  &  F3+ V  &  $0.844\pm0.009$  &  4.490  &  $4.020\pm0.027$  &  --  &  --  & 2, 6 \\
96895  &  G1.5V  &  $0.554\pm0.011$  &  5.990  &  $5.440\pm0.027$  &  $4.731\pm0.050$  &  $4.569\pm0.045$  & 7 \\
98505  &  K2V  &  $0.385\pm0.006$  &  7.670  &  $6.680\pm0.008$  &  --  &  --  & 28 \\
102422  &  K0IV  &  $2.650\pm0.040$  &  3.410  &  $2.510\pm0.027$  &  --  &  $1.201\pm0.051$  & 29 \\
108870  &  K5V  &  $1.881\pm0.017$  &  4.690  &  $3.530\pm0.027$  &  --  &  --  & 10 \\
112447  &  F6V  &  $1.091\pm0.008$  &  4.200  &  $3.590\pm0.027$  &  --  &  $2.851\pm0.080$  & 2 \\
113368  &  A4V  &  $2.230\pm0.020$  &  1.170  &  $1.090\pm0.027$  &  $1.054\pm0.050$  &  $0.981\pm0.051$  & 19 \\
114570  &  F1V  &  $0.648\pm0.008$  &  4.530  &  $4.160\pm0.027$  &  --  &  --  & 3 \\
114622  &  K3V  &  $1.106\pm0.007$  &  5.570  &  $4.470\pm0.027$  &  $3.400\pm0.050$  &  --  & 1 \\
116771  &  F7V  &  $1.082\pm0.009$  &  4.130  &  $3.520\pm0.027$  &  --  &  $2.731\pm0.080$  & 2 \\
120005  &  K7.0V  &  $0.856\pm0.016$  &  7.700  &  --  &  $4.253\pm0.050$  &  --  & 1 \\
\hline
\end{tabular}

\vspace{-12pt}

\tablecomments{{\bf Angular Diameter References:} (1) \citet{boy12b}, (2) \citet{boy12a}, (3) \citet{mae13}, (4) \citet{boy08}, (5) \citet{bai08}, (6) \citet{lig12}, (7) \citet{boy13a}, (8) \citet{dif04}, (9) \citet{dif07}, (10) \citet{dem09}, (11) \citet{van09a}, (12) \citet{dav86}, (13) \citet{moz03}, (14) \citet{ker03a}, (15) \citet{dav11}, (16) \citet{han74}, (17) \citet{chi12}, (18) \citet{nor01}, (19) \citet{ker04a}, (20) \citet{von11b}, (21) \citet{cre12}, (22) \citet{von14}, (23) \citet{van07}, (24) \citet{the05}, (25) \citet{cia01}, (26) \citet{auf06}, (27) \citet{mon12}, (28) \citet{boy15}, (29) \citet{nor99}. {\bf Color Magnitude References:} \citet{neu69}, \citet{gez99}, \citet{car01}, \citet{kid03}, \citet{kim04}, \citet{koe10}, and \citet{mal14}; see Section \ref{data} for more details.}
}
\end{table*}

\begin{table*}
\centering
\caption{Selected Stellar Properties -- Giants}
\label{table:giants_table}
\resizebox{0.85\textwidth}{!}{
\begin{tabular}{|cccccccc|}
\hline
HIP & Sp. Type & $\theta_{LD}$ (mas) & $V$ & $I_C$ & $H$ & $K$ & $\theta_{LD}$ Ref. \\
\hline
3092  &  K3III  &  $4.168\pm0.047$  &  3.270  &  $2.040\pm0.027$  &  $0.551\pm0.050$  &  $0.421\pm0.051$  & 1, 2 \\
7607  &  K3- III CN0.5  &  $3.760\pm0.070$  &  3.590  &  $2.310\pm0.027$  &  --  &  $0.771\pm0.041$  & 3 \\
7884  &  K2/3 III  &  $2.810\pm0.030$  &  4.450  &  $3.050\pm0.027$  &  $1.409\pm0.050$  &  $1.241\pm0.031$  & 3 \\
9884  &  K1IIIb  &  $6.847\pm0.071$  &  2.010  &  $0.860\pm0.027$  &  $-0.558\pm0.050$  &  $-0.649\pm0.051$  & 1, 2, 3, 4 \\
13328  &  K5.5III  &  $4.060\pm0.040$  &  4.560  &  $2.820\pm0.027$  &  --  &  $0.721\pm0.051$  & 1 \\
20205  &  G9.5IIIab CN0.5  &  $2.520\pm0.030$  &  3.650  &  --  &  $1.500\pm0.050$  &  $1.481\pm0.041$  & 5 \\
20455  &  G9.5III CN0.5  &  $2.302\pm0.040$  &  3.770  &  --  &  --  &  $1.581\pm0.051$  & 1, 2, 5 \\
20885  &  G9III Fe-0.5  &  $2.310\pm0.040$  &  3.840  &  --  &  --  &  $1.621\pm0.060$  & 5 \\
20889  &  G9.5III CN0.5  &  $2.572\pm0.046$  &  3.530  &  --  &  --  &  $1.291\pm0.051$  & 1, 2, 5, 6 \\
21421  &  K5III  &  $20.297\pm0.384$  &  0.870  &  --  &  $-2.653\pm0.050$  &  --  & 1, 4, 7, 8 \\
22453  &  K3+ III  &  $2.727\pm0.013$  &  4.890  &  --  &  --  &  $1.441\pm0.041$  & 9 \\
37826  &  G9III  &  $8.177\pm0.130$  &  1.160  &  $0.160\pm0.027$  &  $-1.003\pm0.050$  &  $-1.139\pm0.051$  & 1, 10, 11, 2, 4 \\
42527  &  K1+ III  &  $2.225\pm0.020$  &  4.590  &  $3.420\pm0.027$  &  $1.941\pm0.009$  &  $1.901\pm0.070$  & 12 \\
45860  &  K6III  &  $8.025\pm0.142$  &  3.140  &  $1.460\pm0.027$  &  $-0.475\pm0.050$  &  $-0.699\pm0.031$  & 1, 11, 13, 2 \\
46390  &  K3IIIa  &  $9.700\pm0.100$  &  1.990  &  $0.550\pm0.027$  &  $-1.074\pm0.050$  &  $-1.379\pm0.060$  & 1 \\
49637  &  K3.5IIIb Fe-1:  &  $3.330\pm0.040$  &  4.390  &  $2.890\pm0.027$  &  $1.190\pm0.050$  &  $1.011\pm0.070$  & 3 \\
53229  &  K0+ III-IV  &  $2.540\pm0.030$  &  3.790  &  $2.770\pm0.027$  &  --  &  $1.351\pm0.041$  & 3 \\
54539  &  K1III  &  $4.107\pm0.053$  &  3.000  &  $1.920\pm0.027$  &  $0.539\pm0.010$  &  $0.371\pm0.041$  & 1, 2 \\
55219  &  K0IV  &  $4.745\pm0.060$  &  3.490  &  $2.110\pm0.027$  &  $0.415\pm0.010$  &  $0.251\pm0.041$  & 1, 2 \\
56343  &  G7III  &  $2.386\pm0.021$  &  3.540  &  $2.620\pm0.027$  &  $1.577\pm0.050$  &  $1.491\pm0.060$  & 14 \\
57399  &  K0.5IIIb:  &  $3.230\pm0.020$  &  3.690  &  $2.580\pm0.027$  &  $1.020\pm0.050$  &  $0.901\pm0.031$  & 3 \\
57477  &  K2.5IIIb CN1  &  $1.606\pm0.006$  &  5.270  &  --  &  --  &  $2.531\pm0.060$  & 12 \\
59746  &  K2III  &  $1.498\pm0.028$  &  5.720  &  --  &  --  &  $2.921\pm0.080$  & 12 \\
60202  &  K0III  &  $1.651\pm0.016$  &  4.720  &  $3.720\pm0.027$  &  --  &  $2.321\pm0.051$  & 15 \\
63608  &  G8III  &  $3.254\pm0.037$  &  2.850  &  $1.970\pm0.027$  &  $0.770\pm0.050$  &  $0.731\pm0.051$  & 1, 2, 3 \\
67459  &  K5.5III  &  $4.720\pm0.050$  &  4.050  &  $2.440\pm0.027$  &  --  &  $0.391\pm0.041$  & 3 \\
68594  &  G8:III: Fe-5  &  $0.948\pm0.012$  &  6.180  &  --  &  $3.775\pm0.050$  &  $3.666\pm0.015$  & 16 \\
69673  &  K0III CH-1 CN-0.5  &  $20.877\pm0.277$  &  -0.050  &  $-1.330\pm0.027$  &  $-2.951\pm0.050$  &  --  & 1, 11, 17, 18, 19 \\
72607  &  K4- III  &  $10.300\pm0.100$  &  2.070  &  $0.590\pm0.027$  &  --  &  $-1.259\pm0.070$  & 1 \\
74666  &  G8IV  &  $2.744\pm0.036$  &  3.460  &  $2.480\pm0.027$  &  $1.260\pm0.050$  &  $1.121\pm0.031$  & 1, 2, 3, 6 \\
74793  &  K4III  &  $2.336\pm0.020$  &  5.020  &  --  &  --  &  $1.901\pm0.051$  & 12 \\
75260  &  K4III  &  $1.690\pm0.031$  &  5.720  &  --  &  --  &  $2.721\pm0.060$  & 12 \\
75458  &  K2III  &  $3.596\pm0.015$  &  3.290  &  $2.200\pm0.027$  &  --  &  $0.701\pm0.041$  & 20 \\
77070  &  K2III  &  $4.828\pm0.062$  &  2.630  &  $1.560\pm0.027$  &  $0.197\pm0.007$  &  $0.041\pm0.051$  & 1, 2 \\
79882  &  G9.5IIIb Fe-0.5  &  $2.961\pm0.007$  &  3.230  &  $2.290\pm0.027$  &  --  &  $0.961\pm0.051$  & 21 \\
80331  &  G8III-IV  &  $3.633\pm0.066$  &  2.730  &  $1.890\pm0.027$  &  --  &  $0.601\pm0.031$  & 1, 2 \\
80816  &  G7IIIa Fe-0.5  &  $3.492\pm0.050$  &  2.780  &  $1.880\pm0.027$  &  $0.690\pm0.050$  &  $0.621\pm0.041$  & 1, 2 \\
81833  &  G7III Fe-1  &  $2.529\pm0.050$  &  3.480  &  $2.580\pm0.027$  &  --  &  $1.281\pm0.031$  & 1, 2, 3 \\
82611  &  K2III  &  $1.440\pm0.004$  &  5.990  &  --  &  --  &  $2.811\pm0.090$  & 12 \\
86182  &  K1III  &  $1.515\pm0.010$  &  5.350  &  --  &  --  &  $2.651\pm0.070$  & 12 \\
87833  &  K5III  &  $9.978\pm0.180$  &  2.240  &  $0.630\pm0.027$  &  $-1.160\pm0.050$  &  $-1.319\pm0.041$  & 1, 11, 13, 17 \\
90344  &  K1.5III Fe-1  &  $2.120\pm0.020$  &  4.820  &  $3.630\pm0.027$  &  --  &  $1.931\pm0.051$  & 22 \\
93194  &  A1III  &  $0.753\pm0.009$  &  3.250  &  $3.260\pm0.027$  &  $3.195\pm0.050$  &  --  & 23 \\
94376  &  G9III  &  $3.268\pm0.054$  &  3.070  &  $2.120\pm0.027$  &  --  &  $0.741\pm0.051$  & 1, 2 \\
96837  &  K0III  &  $1.765\pm0.012$  &  4.390  &  $3.410\pm0.027$  &  $2.210\pm0.050$  &  $2.071\pm0.060$  & 9 \\
97938  &  G9.5IIIb  &  $1.726\pm0.008$  &  4.710  &  $3.630\pm0.027$  &  --  &  $2.351\pm0.090$  & 24 \\
98337  &  M0- III  &  $6.821\pm0.098$  &  3.510  &  $1.790\pm0.027$  &  $-0.042\pm0.050$  &  $-0.309\pm0.041$  & 1, 13, 25 \\
99663  &  K5III  &  $1.859\pm0.003$  &  5.810  &  --  &  --  &  $2.311\pm0.070$  & 12 \\
102488  &  K0III-IV  &  $4.610\pm0.050$  &  2.480  &  $1.450\pm0.027$  &  $0.206\pm0.050$  &  $0.101\pm0.070$  & 1 \\
104732  &  G8+ IIIa Ba0.5  &  $2.820\pm0.030$  &  3.210  &  $2.270\pm0.027$  &  $1.155\pm0.050$  &  $1.051\pm0.031$  & 1 \\
110538  &  G9IIIb Ca1  &  $1.920\pm0.020$  &  4.420  &  $3.400\pm0.027$  &  $2.217\pm0.050$  &  $1.961\pm0.051$  & 3 \\
111944  &  K2.5III  &  $2.731\pm0.024$  &  4.500  &  $3.190\pm0.027$  &  $1.600\pm0.007$  &  $1.391\pm0.070$  & 12 \\
\hline
\end{tabular}

\vspace{-12pt}

\tablecomments{{\bf Angular Diameter References:} (1) \citet{moz03}, (2) \citet{nor01}, (3) \citet{nor99}, (4) \citet{1991AJ....101.2207M}, (5) \citet{boy09a}, (6) \citet{van99b}, (7) \citet{2005A&A...433..305R}, (8) \citet{whi87}, (9) \citet{van09a}, (10) \citet{1988ApJ...327..905S}, (11) \citet{dib93}, (12) \citet{bai10}, (13) \citet{1989ApJ...340.1103H}, (14) \citet{the05}, (15) \citet{von14}, (16) \citet{cre12}, (17) \citet{dyc96}, (18) \citet{1996A&A...312..160Q}, (19) \citet{1986A&A...166..204D}, (20) \citet{bai11b}, (21) \citet{maz09}, (22) \citet{lig12}, (23) \citet{mae13}, (24) \citet{bai09}, (25) \citet{2001A&A...377..981W}. {\bf Color Magnitude References:} \citet{neu69}, \citet{gez99}, \citet{car01}, \citet{kid03}, \citet{kim04}, \citet{koe10}, and \citet{mal14}; see Section \ref{data} for more details.}
}

\end{table*}

\subsection{Stellar Classification}
\label{spectral}

We restrict included stars to an effective temperature range of $3900 < T < 10500$ K, which approximately captures spectral types A--K.

Evolved stars are selected not based on their listed spectral classes, but by a stellar radius cut of $6 < R_\star/R_\odot < 100$. This is done in an attempt to disambiguate the luminosity classes of stars which might have inconsistent classifications in the boundary between subgiants and giants.

Some stars in our sample are known to be in multiple star systems. The presence of additional stars can introduce an offset in flux and visibility of the target star. We adopt the selection precedent from \citet{boy08}; we exclude any binary systems where a secondary star is both separated from the target by at most 5$''$ and is within 3 magnitudes of the target in any bands used in the analysis.

\subsection{Angular Diameters}
\label{angdiam}

All stars are required to have limb-darkened angular diameters with mean random errors $\leq$ 2\%, and must have been observed on at least two separate occasions. The measurements come from a variety of sources, which are detailed in \citet{boy12b} and \citet{boy13a} and listed for reference in Tables \ref{table:dwarfs_table}--\ref{table:giants_table}. Stars with inconsistent diameters (here defined as any 2 sources differing by at least 3 times the maximum uncertainty of any measurement) were excluded. We take the uncertainty-weighted means of the remaining measurements for our quoted angular diameters.

Angular diameter source instruments include the Palomar Testbed Interferometer, the Very Large Telescope Interferometer, the Sydney University Stellar Interferometer, the Narrabri Stellar Intensity Interferometer, the Mark III interferometer, the Navy Prototype Optical Interferometer, and especially the CHARA Array.

\subsection{Magnitudes}
\label{mags} 

2MASS photometry \citep{cut03} is saturated for most of the stars in our sample due to brightness. Therefore we rely on earlier photometric catalogs for reliable magnitudes. All magnitudes used are listed in Tables \ref{table:dwarfs_table}--\ref{table:giants_table}. For the $I$ magnitudes we use Cousins $I_C$ photometry converted from Johnson $I_J$ sources \citep{mal14}, as well as magnitudes from \citet{koe10} for our reddest stars. We use \citet{gez99} for $H$ magnitudes, querying the catalog for all magnitude measurements centered at 1.65 $\mu$m. Here, errors of 0.05 mag assumed as a conservative estimate. The $K$ magnitudes are taken from a combination of \citet{neu69}, \citet{kid03}, and \citet{kim04}. Since the filter profiles for the magnitudes in these catalogs differ appreciably, we choose to convert all into the 2MASS system. \citet{neu69} magnitudes are originally in the CIT system \citep{iye82}, and the \citet{kim04} are listed as DENIS $K_S$ magnitudes. \citet{car01} provide transformations from both the CIT and DENIS systems into the 2MASS system (with updated transformations available on the 2MASS website). For the \citet{kid03} magnitudes we first transform into an intermediate system, the Koornneef system \citep{koo83}. \citet{kid03} compare their photometry to the system described in \citep{koo83} and find a constant offset in magnitude. From this we convert to 2MASS via the relation in \citet{car01}. While significant color dependence exists for the DENIS and Koornneef transformations, the CIT transformation exhibits only a very weak color dependence. In light of this, and the lack of accompanying $J$ magnitudes for the \citet{neu69} $K$ magnitudes, we choose to neglect the color term, incorporating the error from this omission into our final uncertainty propagation.

In order to calculate color indices, all stars must have at least one available magnitude in any of the $I_C$, $H$, and $K$ bands. We exclude stars with inconsistent magnitudes: that is, magnitudes from different sources whose values disagree by at least triple the largest uncertainty of any one value.  The listed uncertainties in the resulting colors are propagated from both the uncertainties from conversion as well as assumed 0.02 mag errors in the original Johnson $V$ magnitudes \citep{mal14}.

\section{Fitting Procedure}
\label{fitting}

We choose to construct relations for $V - I_C$, $V - H$, $V - K$, $I_C - H$, and $I_C - K$ (Figure \ref{fig:plots}). We start with a constant-only fit of $\log \theta_{Q=0}$ and add polynomial terms in color, following the form of Equation \ref{eq:polyfit}. The fitting procedure uses a Levenberg-Markquardt least-squares algorithm provided by the {\it MPFIT} routine \citep{mar09}. For each solution, we perform an F-test \citep{pre92} to determine whether the improvement to the relation by adding a polynomial term is statistically significant. Once the functional form is obtained we run a Monte Carlo simulation by generating $10^4$ simulated datasets, randomly choosing colors and diameters drawn from Gaussian probability distributions of each star's true color and diameter. The means and standard deviations are given by the initial fit coefficients and their associated uncertainties, respectively. This allows us to incorporate all measurement uncertainties into the relation.

\section{Analysis}
\label{analysis}

\begin{figure*}
\begin{center}
   \begin{tabular}{cc}
\includegraphics[width=7.5cm]{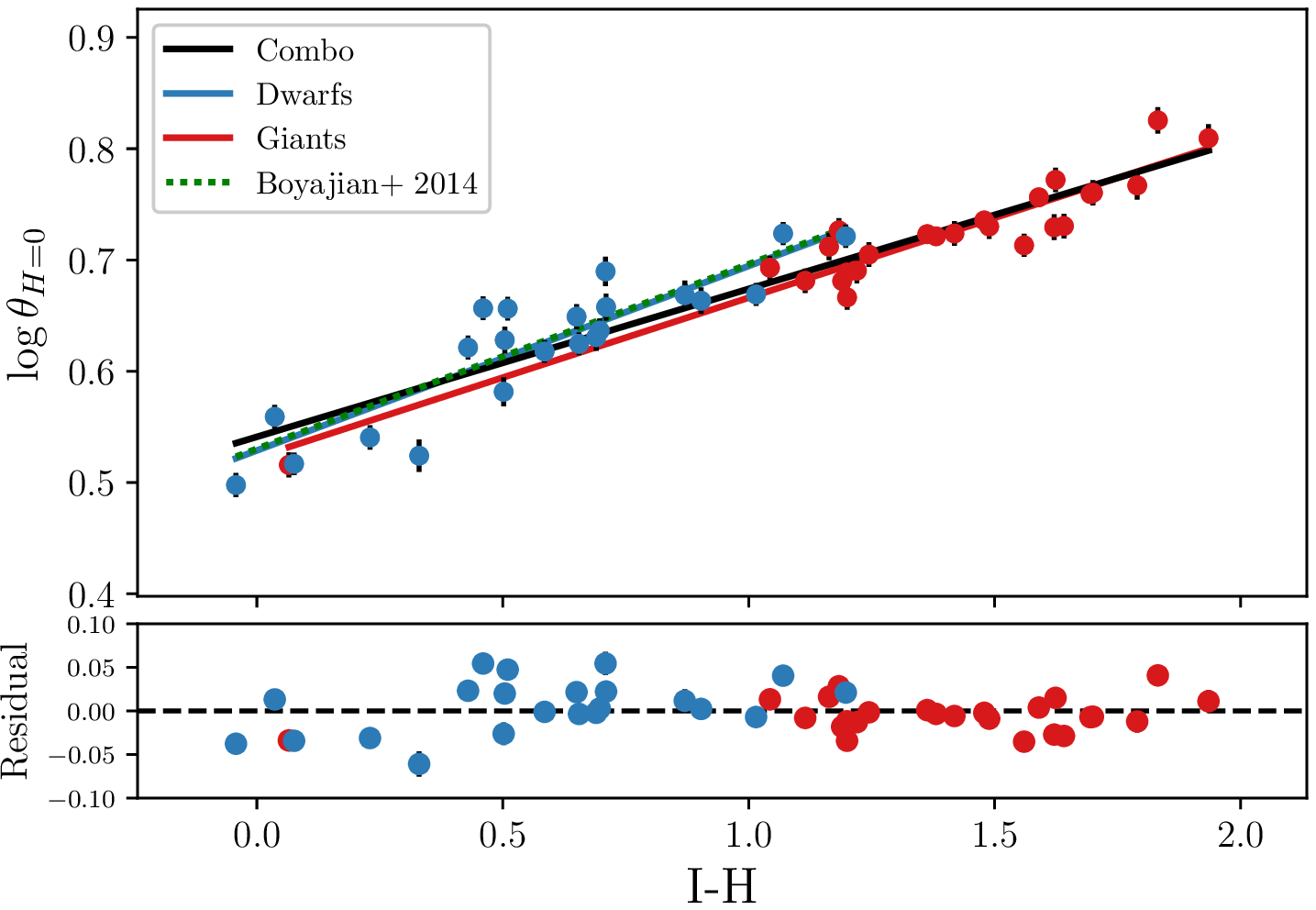} &
\includegraphics[width=7.5cm]{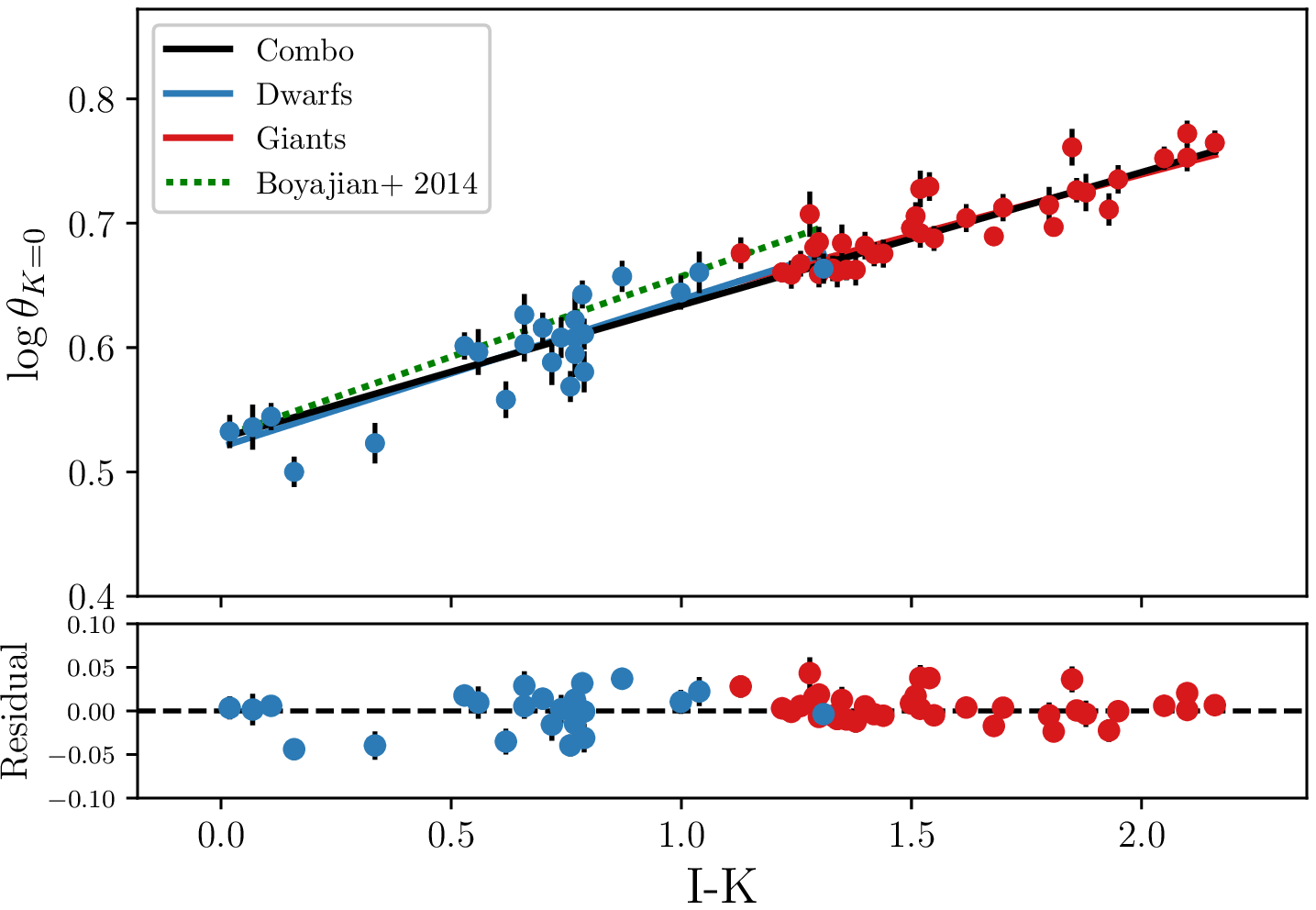} \\
\includegraphics[width=7.5cm]{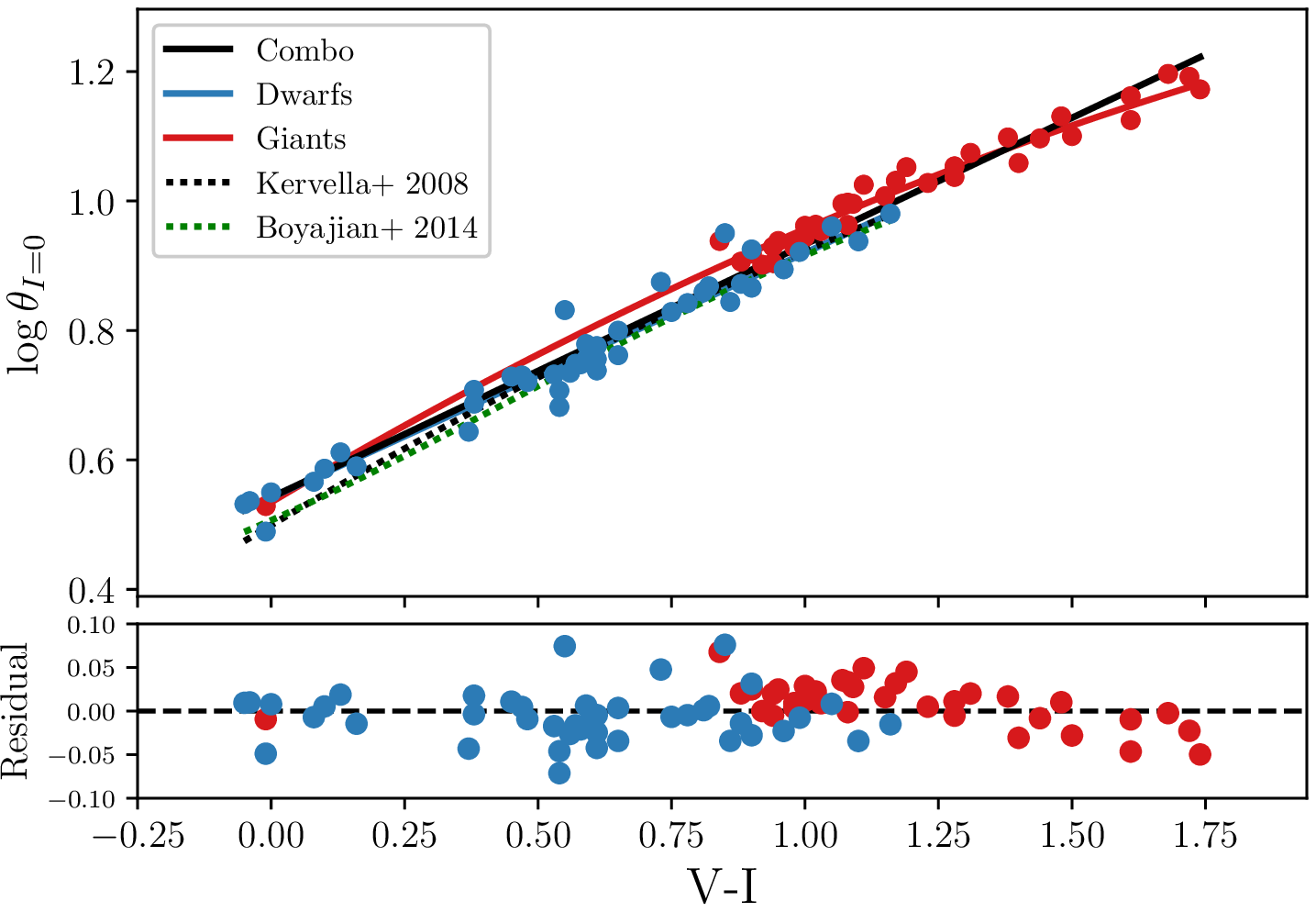} &
\includegraphics[width=7.5cm]{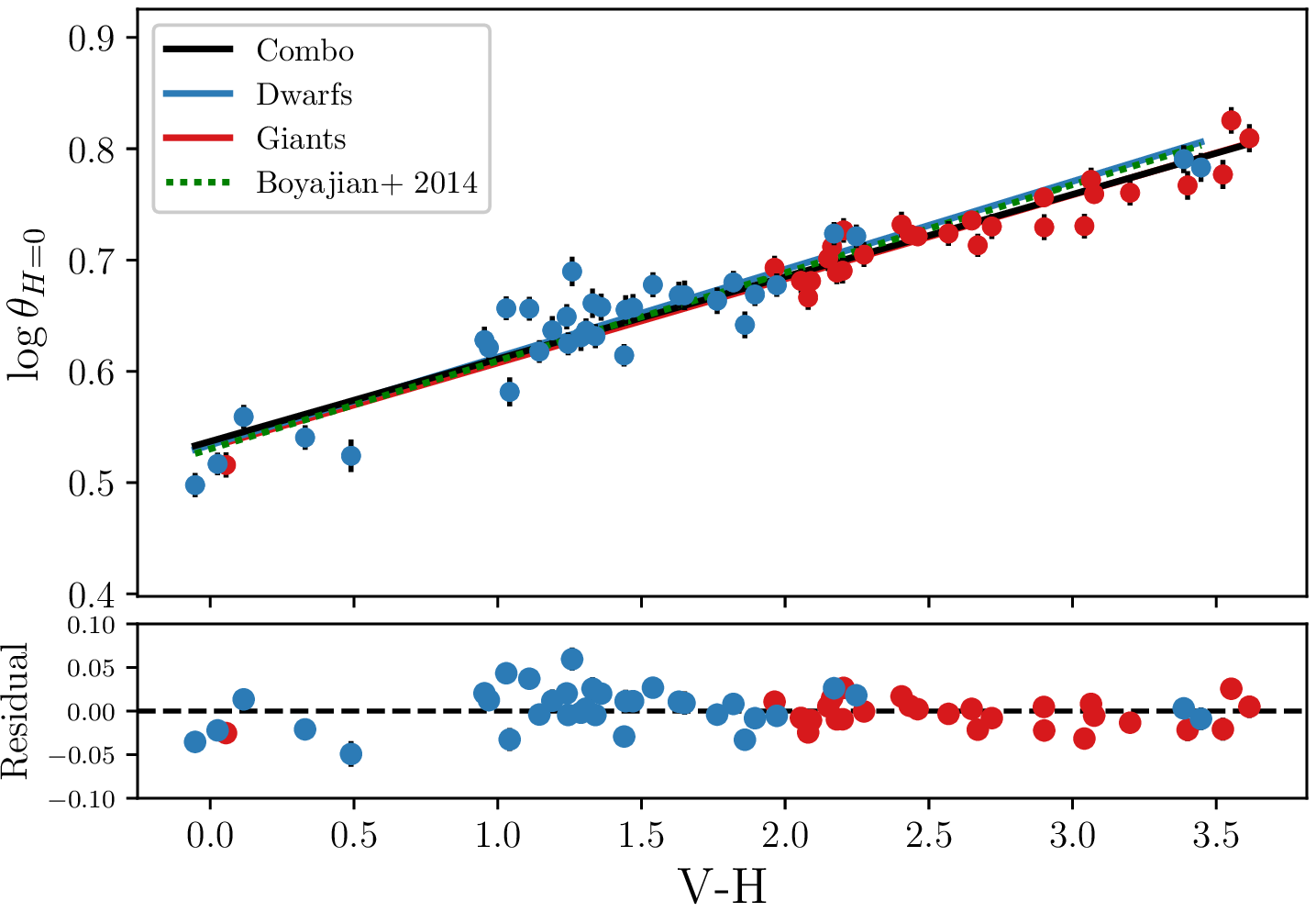} \\
\includegraphics[width=7.5cm]{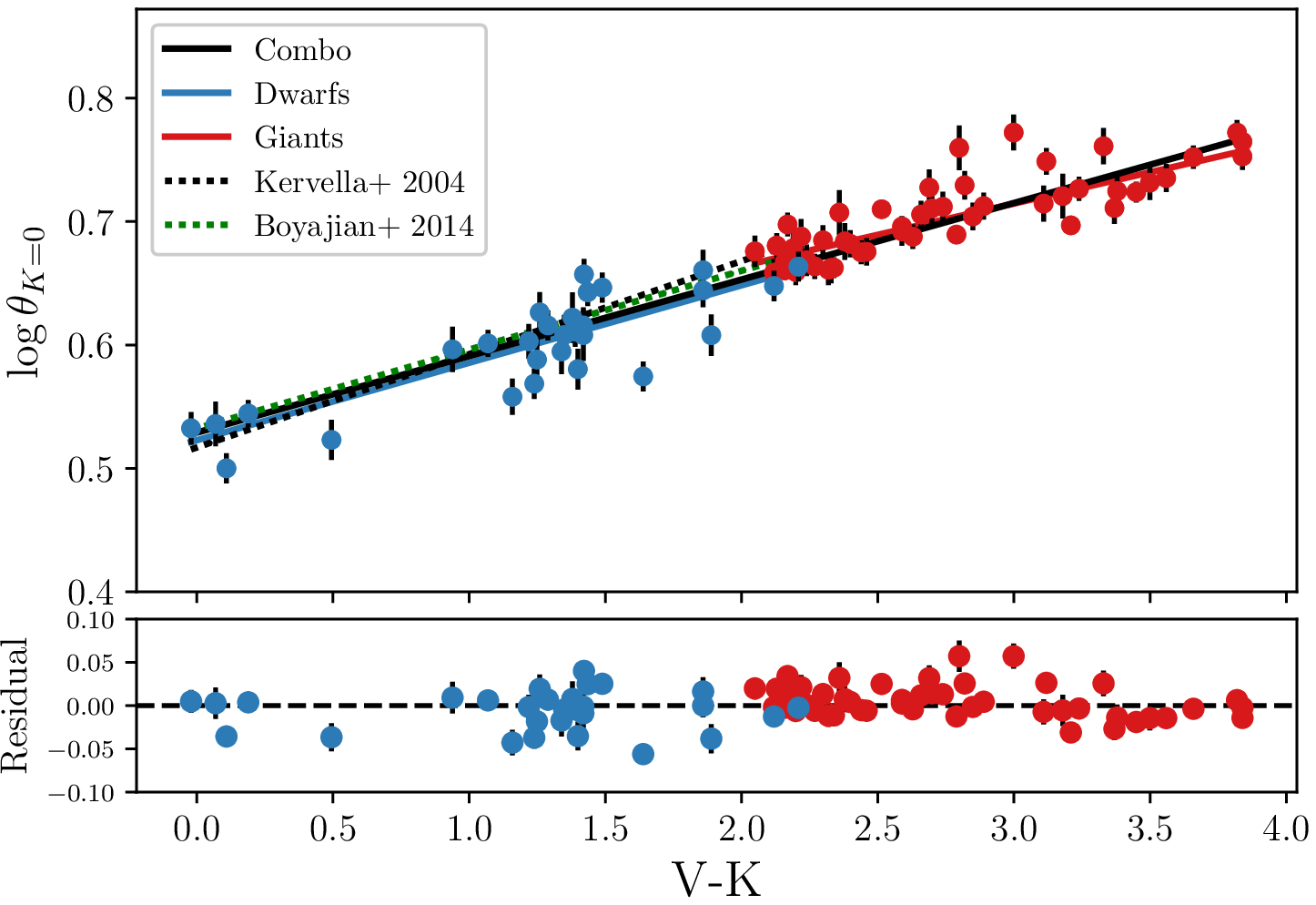}
    \end{tabular}
\caption{{\it Top panels:} Angular diameter-color relations for both dwarf/subgiants (blue line) and giants (red line), as well as a combined fit (black line). The functional form of the fits is described in Equation \ref{eq:polyfit} with coefficients listed in Table \ref{table:relations}. The data are introduced in Section \ref{data} and catalogued in Tables \ref{table:dwarfs_table}--\ref{table:giants_table}. The fitting methodology is described in Section \ref{analysis}. All panels show previous relations from \citet{boy14}. For $V-I_C$ we include the result for dwarfs from \citet{ker08a}, and for $V-K$ we include the result for dwarfs from \citet{ker04c}. {\it Bottom panels:} The residuals in dex are shown with respect to the combined relation.}
\label{fig:plots}
\end{center}
\end{figure*}

For each color we construct independent fits of unevolved and evolved stars (Figure \ref{fig:plots}). Table \ref{table:relations} shows the number of stars, range of colors, and fit coefficients for each fit. These relations are only valid for the color ranges for which we have data. We then construct fits for each color using all stars, unevolved and evolved. The combined fits test that variations in surface gravity with stellar evolution will not affect the relations (as noted in Section \ref{introduction}). The derived relations with the combined sample have similar RMS errors to the separated fits, which is consistent with the result of \citet{bar76} that surface brightness is independent of luminosity class.  The RMS in the residuals ranges from 0.017 to 0.03 dex, corresponding to the minimum expected uncertainties in $\log \theta_{\mathrm{LD}}$ before uncertainties in magnitudes are considered. The $I_C - K$ relations have the smallest spread for all fits.

To estimate the uncertainties in limb-darkened diameters, we propagate uncertainties for assumed 0.03 errors in both magnitudes of a given color. The most precise relations for all stars are those for $V - H$ and $V - K$, which have estimated uncertainties of 1.8--2.9\%. (The range is due to the dependence of the uncertainty on the color, which we have varied within the range of the sample.) The least precise results are in $V - I_C$, where the corresponding uncertainties could be as high as 6.5\%.

\begin{table*}
\centering
\caption{Angular Diameter - Color Relations}
\label{table:relations}
\begin{tabular}{|cccccccc|}
\hline
Color & $N$ & Range & $c_0$ & $c_1$ & $c_2$ & RMS (dex) & Pred. Frac. Uncertainty \\
\hline
{\bf All Stars} \\
$I_C-H$ & 47 & -0.042--1.935 & 0.541$\pm$0.004 & 0.133$\pm$0.003 & --- & 0.025 & 0.021--0.031 \\
$I_C-K$ & 60 & 0.019--2.159 & 0.528$\pm$0.005 & 0.108$\pm$0.003 & --- & 0.020 & 0.021--0.031 \\
$V-I_C$ & 83 & -0.050--1.740 & 0.542$\pm$0.006 & 0.391$\pm$0.006 & --- & 0.028 & 0.043--0.065\\
$V-H$ & 63 & -0.052--3.615 & 0.538$\pm$0.004 & 0.074$\pm$0.002 & --- & 0.020 & 0.018--0.029 \\
$V-K$ & 78 & -0.021--3.839 & 0.529$\pm$0.004 & 0.062$\pm$0.002 & --- & 0.021 & 0.018--0.029 \\
{\bf Dwarfs and Subgiants} \\
$I_C-H$ & 22 & -0.042--1.198 & 0.529$\pm$0.007 & 0.166$\pm$0.010 & --- & 0.031 & 0.026--0.049 \\
$I_C-K$ & 24 & 0.019--1.309 & 0.520$\pm$0.007 & 0.118$\pm$0.010 & --- & 0.023 & 0.024--0.047 \\
$V-I_C$ & 45 & -0.050--1.160 & 0.542$\pm$0.007 & 0.378$\pm$0.011 & --- & 0.029 & 0.043--0.070 \\
$V-H$ & 35 & -0.052--3.447 & 0.534$\pm$0.005 & 0.079$\pm$0.003 & --- & 0.023 & 0.020--0.036 \\
$V-K$ & 29 & -0.021--2.209 & 0.523$\pm$0.006 & 0.063$\pm$0.005 & --- & 0.024 & 0.022--0.038 \\
{\bf Giants} \\
$I_C-H$ & 25 & 0.065--1.935 & 0.523$\pm$0.011 & 0.144$\pm$0.007 & --- & 0.020 & 0.033--0.054\\
$I_C-K$ & 36 & 1.129--2.159 & 0.543$\pm$0.011 & 0.098$\pm$0.007 & --- & 0.016 & 0.040--0.053\\
$V-I_C$ & 38 & -0.010--1.740 & 0.535$\pm$0.027 & 0.490$\pm$0.046 & -0.068$\pm$0.019 & 0.026 & 0.080--0.241\\
$V-H$ & 28 & 0.055--3.615 & 0.532$\pm$0.009 & 0.076$\pm$0.003 & --- & 0.016 & 0.026--0.041\\
$V-K$ & 49 & 2.049--3.839 & 0.562$\pm$0.009 & 0.051$\pm$0.003 & --- & 0.019 & 0.031--0.040\\
\hline
\end{tabular}
\vspace{-12pt}

\tablecomments{Numerical values for the relations in Figure \ref{fig:plots}. The color index, number of stars per index, range of color, and fit coefficients for Equation \ref{eq:polyfit} (which takes the form $\log \theta_{Q=0} = \sum_{n=0}^N c_n \left(P - Q\right)^n$) are shown. For each relation we have calculated both the RMS of the relation and the range of propagated fractional uncertainties for each zero-magnitude angular diameter, assuming 0.03 mag errors in each band.}

\end{table*}

Initially M dwarfs were included in our sample, to see if the derived relation would change drastically with their inclusion. The addition of dwarfs at $T<3900$ K adds a statistically significant quadratic coefficient to our fits in the $V - I_C$ relation. Our temperature cut therefore provides a more precise relation for FGK stars in particular. In contrast the $V - I_C$ relation in giants has a marginally significant quadratic term, even excluding stars below 3900 K. On the other end of our temperature range, inclusion of the A dwarfs (and one A giant) did not significantly change the fits, and so we are less hesitant to include them here.

We also test whether the angular diameter relations have a statistically significant dependence on stellar metallicity. Metallicity has shown to be a factor in relations of stellar radii to color indices \citep{boy12b}, since changes in metallicity tend to affect bluer parts of the stellar spectra due to line blanketing \citep{mcn69}. Such effects would propagate to stellar angular diameter relations, but in all colors considered in this paper the relations were insensitive to metallicity. This is consistent with the findings of \citet{boy14}, which found metallicity in their angular diameter relations was strongest for the colors with the shortest wavelength bands ($B - V$, $g - r$), where the bluer colors would be affected more strongly by line blanketing.

\subsection{Comparison with Previous Works}\label{sub:comp}
We directly compare our relations to those of \citet{boy14}, and for $V - I_C$ to the relation in Table 3 of \citet{ker08a}, and the $V - K$ relation to Eq. (23) of \citet{ker04c}, as seen in Figure \ref{fig:plots}. All the mentioned relations are valid for dwarfs and subgiants, and it should be noted that all extend through the M spectral type (not shown here).  In $V - I_C$ the largest offset in angular diameter between the data and the \citet{ker08a} prediction is 0.08 dex, and for $V - K$ the largest offset with respect to \citet{ker04c} is 0.06 dex. We expect at least reasonable agreement with the results of \citet{boy14} and \citet{ker08a} (see Figure \ref{fig:plots}) by construct, since we include the subset of angular diameters used in these works which meet our uncertainty constraint $\left( \leq 2 \% \right)$. Nevertheless, our sources differ from these works for $I_C$ \citep{koe10,mal14}, $H$ \citep{gez99}, and $K$ \citep{neu69,kid03,kim04} band photometry, as well as a larger sample within the FGK color range. Hence differences in the predicted angular diameters exist, particularly on the blue end of the $V - I_C$ relation, where both \citet{ker08a} and \citet{boy14} use higher-order polynomial fits that underestimate the diameters of the bluest dwarfs, while fitting well for dwarfs well beyond the red end of our color range.

\section{Conclusions}

This work describes new relations linking stellar angular diameter to photometric colors. We use a dataset with roughly twice the precision in angular diameter measurements compared to previous papers. We use empirical evidence that predictions of angular diameters from color and magnitude are insensitive to luminosity class to construct, for the first time, a prediction of angular diameters at fixed magnitude for A--K stars across the stages of stellar evolution. We find that there is no dependence on stellar metallicity for the colors tested.

Further improvement to the relations will require additional angular diameter measurements to fill in parameter space for the earlier-type giants. Additionally, lack of demonstrably consistent $I_C$ photometry for M stars of any luminosity class limits us from extending the red end of our relations. Transformations among systems are susceptible to differences in zero points, susceptibility of filters to red leaks, and correlated errors in the filter profiles \citep{man15}. Nevertheless, for FGK stars the continuity in the relation between color and angular size is well-constrained by the regions of overlap of the spectral classes.

\section*{Acknowledgements}

This work supported by a graduate fellowship from the Gruber Foundation. Additional gratitude to Anthony Mallama for providing converted Cousins I magnitudes for the dwarfs in our sample.

This research has made use of the NASA Exoplanet Archive, which is operated by the California Institute of Technology, under contract with the National Aeronautics and Space Administration under the Exoplanet Exploration Program.

%{\it Facilities:} \facility{CHARA}, \facility{PO:PTI}, \facility{VLTI}, Sydney University Stellar Interferometer, Narrabri Stellar Intensity Interferometer, \facility{Mark III interferometer}, \facility{NPOI}.

%Citation of Python software packages not explicitly mentioned in the paper but used in the analysis.
\nocite{fre01}
\nocite{hun07}
\nocite{mck10}
\nocite{van11}

%%%%%%%%%%%%%%%%%%%%%%%%%%%%%%%%%%%%%%%%%%%%%%%%%%



\bibliographystyle{mnras}
\bibliography{paper} % if your bibtex file is called example.bib

%%%%%%%%%%%%%%%%%%%%%%%%%%%%%%%%%%%%%%%%%%%%%%%%%%

%%%%%%%%%%%%%%%%% APPENDICES %%%%%%%%%%%%%%%%%%%%%

%\appendix


%%%%%%%%%%%%%%%%%%%%%%%%%%%%%%%%%%%%%%%%%%%%%%%%%%

% Don't change these lines
\bsp	% typesetting comment
\label{lastpage}
\end{document}